\documentclass[a4paper,11pt]{article} 
\usepackage[margin=3cm]{geometry}
\usepackage{amstext,amsmath,amssymb}
\usepackage{blindtext}
\usepackage{enumitem}
\usepackage{changepage, environ}
\PassOptionsToPackage{pass,showframe}{geometry}
\usepackage{float}
\usepackage[utf8]{inputenc}
\usepackage[T1]{fontenc}
\usepackage[autostyle]{csquotes}
\usepackage{dirtytalk}
\usepackage{graphicx} 
\usepackage{caption}
\usepackage[english]{babel}
\usepackage{xcolor}
\usepackage[breaklinks]{hyperref}
\usepackage{setspace}
\hypersetup{colorlinks=true, pdfstartview=FitV, linkcolor=blue, citecolor=black, plainpages=false, pdfpagelabels=true, urlcolor=blue}
\usepackage{xurl}
\usepackage{scalerel}
\usepackage{tikz}
\usepackage{multicol}
\usepackage{blindtext}
\usepackage{abstract}
\usepackage{sectsty}
\usepackage{secdot}
\usepackage{fancyhdr}
\usepackage{secdot}
\usepackage{blindtext}
\usepackage{academicons}
\usepackage{graphicx}
\graphicspath{ {./images/} }

\PassOptionsToPackage{dvipsnames}{xcolor}
\usetikzlibrary{decorations.text}
\usepackage{tikz}
\usepackage[nottoc,numbib]{tocbibind}
\usepackage{setspace}

\IfFileExists{upquote.sty}{\usepackage{upquote}}{}
\IfFileExists{microtype.sty}{% use microtype if available
  \usepackage[]{microtype}
  \UseMicrotypeSet[protrusion]{basicmath} % disable protrusion for tt fonts
}{}
\makeatletter
\@ifundefined{KOMAClassName}{% if non-KOMA class
  \IfFileExists{parskip.sty}{%
    \usepackage{parskip}
  }{% else
    \setlength{\parindent}{0pt}
    \setlength{\parskip}{6pt plus 2pt minus 1pt}}
}{% if KOMA class
  \KOMAoptions{parskip=half}}
\makeatother
\usepackage{longtable,booktabs,array}
\usepackage{calc} % for calculating minipage widths

\title{\textbf{Subject Roles in the EU AI Act: Mapping and Regulatory Implications}}

\usepackage[multiple]{footmisc}
\usepackage{bigfoot}

\DeclareNewFootnote{AAffil}[arabic]
\DeclareNewFootnote{ANote}[fnsymbol]

\NewEnviron{myquote}[1][\relax]{
  \begin{center}
  \begin{adjustwidth}{50pt}{50pt}
  \setbox0=\hbox{\itshape\BODY}%
  \begin{itshape}
  \ifdim\wd0>\linewidth\relax\BODY\else\hfil\BODY\fi
  \end{itshape}
  \end{adjustwidth}
  \end{center}  
  \ifx\relax#1\relax\else
    \vspace{-0.15cm}
    \begin{adjustwidth}{0pt}{20pt}
    \flushright { \small -\hspace{2pt}#1\hspace{2pt}- }
    \end{adjustwidth}
    \vspace{0.3cm}
  \fi
}

\author{
  Nicola Fabiano\thanks{\href{https://www.fabiano.law/en/page/about/}{Studio Legale Fabiano (Italy)} - Affiliation: \textit{International Institute of Informatics and Systemics (IIIS) - USA}}
}

\date{September 28, 2025}

\begin{document}
\maketitle

\begin{abstract}
The European Union's Artificial Intelligence Act (Regulation (EU) 2024/1689) establishes the world's first comprehensive regulatory framework for AI systems through a sophisticated ecosystem of interconnected subjects defined in Article 3. This paper provides a structured examination of the six main categories of actors — providers, deployers, authorized representatives, importers, distributors, and product manufacturers — collectively referred to as "operators" within the regulation. Through examination of these Article 3 definitions and their elaboration across the regulation's 113 articles, 180 recitals, and 13 annexes, we map the complete governance structure and analyze how the AI Act regulates these subjects. Our analysis reveals critical transformation mechanisms whereby subjects can assume different roles under specific conditions, particularly through Article 25 provisions ensuring accountability follows control. We identify how obligations cascade through the supply chain via mandatory information flows and cooperation requirements, creating a distributed yet coordinated governance system. The findings demonstrate how the regulation balances innovation with the protection of fundamental rights through risk-based obligations that scale with the capabilities and deployment contexts of AI systems, providing essential guidance for stakeholders implementing the AI Act's requirements.
\end{abstract}

\noindent \textbf{Keywords.} EU AI Act, Article 3, Operators, Providers, Deployers, Role Transformation, Regulatory Obligations, Information Flows, Importers, Distributors, Authorized Representatives

\section{Introduction}\label{introduction}

The European Union's Artificial Intelligence Act, adopted on June 13, 2024, as Regulation (EU) 2024/1689, marks a significant shift in technology governance, establishing comprehensive rules for AI systems within a carefully defined ecosystem of stakeholders. Article 3 of the regulation provides precise definitions for six primary categories of actors: providers, deployers, authorized representatives, importers, distributors, and product manufacturers, collectively designated as "operators" under Article 3(8).

This paper conducts an in-depth examination of these entities, exploring how the AI Act governs them across 113 articles, 180 recitals, and 13 annexes.

The significance of understanding these subject roles as defined in Article 3 extends beyond mere compliance. The regulation's sophisticated approach to defining and regulating these operators reflects a deep understanding of how modern AI systems are developed, deployed, and maintained. The Act acknowledges that AI systems frequently involve multiple contributors across complex value chains, ranging from those developing underlying neural network architectures to organizations deploying them in real-world contexts. Each operator category bears specific obligations calibrated to their role and control over AI systems.

The concept of "operator" itself, introduced in Article 3(8), creates an overarching category encompassing all primary actors in the AI ecosystem. This collective designation enables the regulation to establish common provisions applicable to all actors while maintaining differentiated obligations based on specific roles. The inclusion of "product manufacturer" within the operator definition addresses scenarios where AI systems are embedded in products, ensuring comprehensive coverage of all entities exercising control over AI systems.

This paper systematically analyzes each operator category as defined in Article 3, tracing how these definitions are elaborated and operationalized throughout the regulation. We examine not only the explicit obligations assigned to each operator type but also the complex web of relationships connecting them, the mechanisms for role transformation, and the practical implications for organizations operating within this regulatory framework.

\section{Methodology}\label{methodology}

Our methodological approach involves a systematic analysis of how the six operator categories defined in Article 3 are regulated throughout the AI Act, employing a multi-layered analytical framework that captures both the legal architecture and practical implications of the regulatory system. The research encompasses three interconnected phases: definitional analysis, comprehensive regulatory mapping, and synthetic interpretation of obligations and relationships.

The first phase involves detailed semantic and legal examination of the Article 3 definitions for providers (3(3)), deployers (3(4)), authorized representatives (3(5)), importers (3(6)), distributors (3(7)), and the collective operator concept (3(8)). We analyze the precise language of these definitions using established principles of EU legal interpretation, examining the ordinary meaning of each term, its contextual usage within the regulation, and its teleological purpose within the broader regulatory framework. Particular attention one should pay to definitional boundaries---understanding not only what each definition includes but also what it excludes or leaves ambiguous. For instance, the provider definition's inclusion of entities that "have an AI system developed" requires careful analysis of what level of control or involvement triggers provider status when development is outsourced. Similarly, the deployer definition's exclusion of "personal non-professional activity" necessitates examination of where the boundary lies between professional and personal use, particularly in edge cases such as freelance professionals or small business owners.

The second phase involves comprehensive regulatory mapping through systematic document analysis of the regulation's 113 articles and 13 annexes. Our mapping methodology employs both keyword searching and contextual reading to ensure complete identification of all provisions affecting each operator type. For each operator category, we create detailed provision matrices categorizing references by regulatory type: substantive obligations (what operators must do), procedural requirements (how they must do it), rights and powers (what they may do), relational provisions (how they interact with other operators), and transformation mechanisms (when they assume different roles). This mapping extends beyond explicit mentions to capture implicit applications---for instance, provisions referring to "economic operators" or using passive constructions that imply operator obligations without naming specific actors. We also trace cross-references between articles to understand how obligations in one provision trigger requirements in others, creating regulatory chains that might not be apparent from an isolated reading.

The synthesis phase integrates our findings through a systematic analysis that employs multiple theoretical frameworks. From a regulatory theory perspective, we examine how the operator definitions create a responsive regulatory system that allocates responsibilities based on control and capacity. From a network analysis perspective, we map information flows and interdependencies between operators to understand how the regulation creates governance through interconnection rather than hierarchy alone. From a practical implementation perspective, we consider how organizations must operationalize these definitions in real-world contexts where roles may be fluid or contested. Throughout this analysis, we systematically incorporate the 180 recitals that provide essential interpretive context. The recitals often clarify ambiguities in the operative provisions, explain the rationale for specific operator obligations, and indicate how provisions should be applied in practice. Our methodology treats recitals not as mere preamble but as integral components of the regulatory framework that shape how one should understand operator definitions and obligations.

%%%------------------
\begin{figure}[H]
\centering
\includegraphics[width=0.8\textwidth]{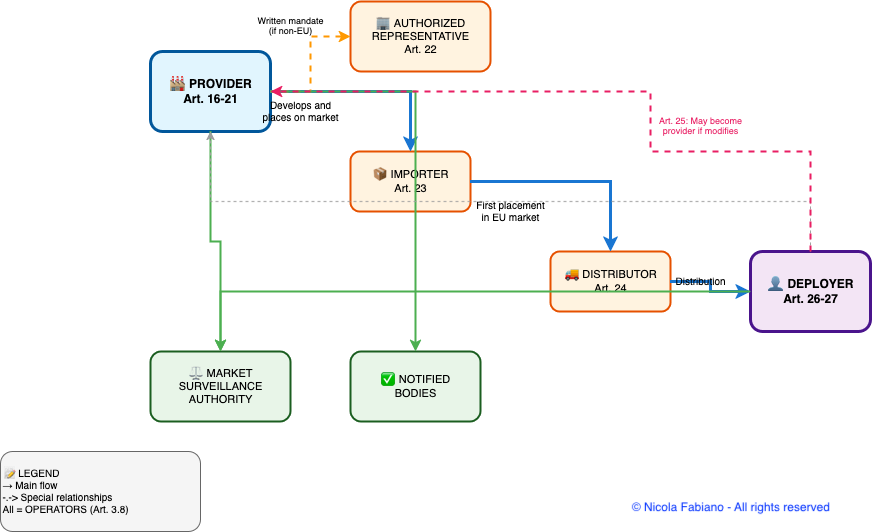}
\caption{Subjective roles according to Article 3 of the EU AI Act}
\label{fig:subjective-roles}
\end{figure}
%%%-----------------

\section{The Provider: Primary Responsibility in the AI Value Chain}\label{provider}

\subsection{Definition and Scope}\label{provider-definition}

Article 3(3) defines the provider as "\textit{a natural or legal person, public authority, agency or other body that develops an AI system or a general-purpose AI model or that has an AI system or a general-purpose AI model developed and places it on the market or puts the AI system into service under its name or trademark, whether for payment or free of charge}".

This definition establishes the provider as the primary accountability point in the AI ecosystem, embodying the regulatory principle that those who exercise decisive control over AI system creation and commercialization must bear primary responsibility for safety and compliance. Crucially, the provider sits at the apex of the operator hierarchy defined in Article 3(8), serving as the central node from which information flows bidirectionally to all other operators in the value chain.

The breadth and sophistication of this definition merit careful analysis. The inclusion of both direct development ("develops") and commissioned development ("has...developed") ensures that entities cannot escape provider obligations through outsourcing arrangements. This captures various business models, from technology companies with in-house AI development teams to enterprises that commission custom AI solutions from third-party developers. The critical factor is not who performs the technical work but who exercises commercial control by placing the system on the market under their identity. This approach prevents regulatory arbitrage and is reinforced by Article 25, which can transform other operators into providers when they make substantial modifications or change the intended purpose of AI systems. Regulatory arbitrage would occur if entities could escape provider obligations by outsourcing technical development to third parties while retaining commercial control and market benefits---a practice this framework explicitly prevents by assigning responsibility based on who places the system on the market under their name, not who codes it.

The definition's scope extends to both "AI systems" and "general-purpose AI models," reflecting the technological reality that modern AI often involves foundation models that serve as platforms for multiple applications. A provider of a large language model bears obligations not only for the model itself but also must consider how it might be integrated into downstream systems. This creates cascading information obligations under Article 13, requiring providers to furnish comprehensive documentation to deployers, and under Article 53 for general-purpose AI models, mandating the provision of information to downstream providers integrating these models. The phrase "\textit{whether for payment or free of charge}" ensures that open-source AI providers and those offering free services remain within regulatory scope, recognizing that market impact and potential risks do not depend on monetization models. The reference to placement "under its name or trademark" establishes the decisive criterion: public presentation as the source of the AI system, regardless of underlying development arrangements.

\subsection{Provider Obligations Throughout the EU AI Act}\label{provider-obligations}

The provider's obligations extend far beyond the basic definition, creating a comprehensive compliance framework that touches every aspect of AI system development, deployment, and lifecycle management. As the primary operator category, providers bear the most extensive set of obligations, appearing in over 80 articles throughout the regulation. Articles 16-21 establish the foundational obligations for providers of high-risk AI systems, creating a systematic approach to safety and compliance that begins before development and extends throughout the operational life of these systems.

Article 16 serves as the cornerstone obligation, mandating that providers ensure their high-risk AI systems comply with all requirements set out in Chapter 2 of the regulation. This seemingly simple requirement encompasses a vast array of specific obligations, including risk management systems, data governance, technical documentation, transparency, human oversight, accuracy, robustness, and cybersecurity. The provision makes clear that compliance is not optional or partial---providers must ensure their systems meet all applicable requirements before placing them on the market or putting them into service. This creates a fundamental duty that ripples through the entire operator chain, as Article 23 requires importers to verify provider compliance, Article 24 mandates distributors to verify documentation, and Article 26 obligates deployers to use systems by the provider's instructions.

Article 17's requirement for quality management systems represents a paradigm shift in how AI development must be organized. Unlike traditional software development, where quality processes may be informal or ad hoc, the regulation requires systematic and documented approaches that are proportionate to an organization's size and AI activities. This system must interface with other operators in accordance with Article 25(4), which requires written agreements to specify the exchange of information when third parties supply tools, services, or components integrated into high-risk AI systems. For large providers developing multiple AI systems, this means enterprise-wide governance frameworks with clear policies, procedures, and accountability structures. For smaller providers, it requires, at a minimum, documented processes for risk assessment, design control, testing, and post-market surveillance. The quality management system must address the entire AI lifecycle, from initial conception through retirement, ensuring that compliance is built into organizational DNA rather than bolted on as an afterthought.

Article 18's technical documentation requirements create unprecedented transparency obligations for AI systems, particularly those based on complex neural networks. This documentation forms the backbone of the information flow to other operators: importers must verify its existence (Article 23(1)(b)), distributors must ensure its availability (Article 24(1)), and deployers must reference it for proper system use (Article 26(9)). The provider must prepare and maintain up-to-date documentation before placing it on the market and throughout the system's lifecycle. Annex IV specifies extensive documentation requirements, including a general system description, a detailed description of AI model development (including training methodologies and datasets), information about system architecture (including computational logic and algorithms), and a description of system testing and validation procedures. For neural network-based systems, this requires documenting not only the final architectures but also the experimental process that led to design decisions, including failed approaches and the rationale behind specific choices. Organizations must develop new documentation practices that can capture the iterative and experimental nature of AI development while maintaining clarity for regulatory review.

Article 19 mandates that high-risk AI systems be designed with capabilities for automatic recording of events ("logs") throughout their lifetime. These logs become a critical tool for operator cooperation: providers must maintain them (Article 19), deployers must retain them when under their control (Article 26(6)), and both must provide them to authorities during investigations (Article 73). This requirement ensures traceability of system behavior and enables investigation of incidents or malfunctions. For providers, this means incorporating logging capabilities into the system architecture from the outset, rather than adding them as an afterthought. The logs must capture sufficient detail to understand system operation while respecting data protection requirements---a delicate balance requiring careful design. The specific logging requirements vary by AI system type, but generally require the ability to understand when the system was used, with what input data, what decisions were made, and any anomalies or errors that occurred.

Article 20 establishes a dynamic compliance obligation with crucial multi-operator dimensions. When providers identify non-compliance, they must immediately inform distributors and, where applicable, deployers (Article 20(1)). They must investigate causes "in collaboration with the reporting deployer" (Article 20(2)), creating mandatory cooperation loops. Providers are required to establish procedures to detect instances of non-compliance, whether uncovered during testing, reported by users, or identified by regulators. Once such issues are detected, they must be promptly investigated, the underlying causes identified, and corrective measures taken. If the system presents risks, providers must consider withdrawal or recall, balancing commercial interests against safety obligations. The requirement to inform distributors and authorities ensures transparency and enables a coordinated response across the supply chain.

Article 21 codifies cooperation obligations with competent authorities, requiring providers to provide all information and documentation necessary to demonstrate conformity. This cooperation extends through the operator chain, as Article 79(2) requires all "relevant operators" to cooperate with authorities during risk investigations, and Article 82 mandates coordinated responses when compliant systems present risks. Providers must respond to information requests promptly and comprehensively, provide access to AI systems for testing when requested, and cooperate with market surveillance activities. The cooperation requirement entails providing authorities with access to automatically generated logs and, when necessary for conformity assessments, making the training datasets used by deep learning-based AI systems available.

For providers of general-purpose AI models, Articles 53-55 establish additional specialized obligations recognizing the unique challenges these systems pose. Article 53(2) specifically requires providers to provide documentation to downstream providers integrating their models, creating a technical information cascade through the AI value chain. Article 53 applies to all general-purpose AI model providers, requiring them to develop and maintain up-to-date technical documentation of the model, which includes, at a minimum, the information set out in Annex XI. This documentation must capture the complexity of models that may contain billions of parameters organized in sophisticated neural network architectures. Providers must also provide information and documentation to downstream providers who integrate their models into AI systems, enabling safe and compliant integration. The copyright compliance requirement addresses one of the most contentious aspects of large model training — the use of vast datasets that may include copyrighted material. Providers must implement policies respecting Union copyright law, including mechanisms to honor rights-holder opt-outs under Directive 2019/790.

Article 55 imposes additional obligations on general-purpose AI models with systemic risks, primarily defined by a computational threshold of $10^{25}$ FLOPs for training. These models must undergo model evaluation, including adversarial testing, to identify potential dangerous capabilities or misuse vectors. Providers must assess and mitigate systemic risks that could arise from model deployment at scale. Article 60(2) enables testing "in partnership with one or more deployers," establishing formal provider-deployer collaboration mechanisms for real-world testing. These obligations acknowledge that the most powerful AI models pose qualitatively distinct risks that require enhanced governance.

The provider's responsibilities extend through post-market monitoring under Article 72, establishing obligations that continue throughout the AI system's operational life. Article 72(2) explicitly states that monitoring systems must collect "data which may be provided by deployers," formalizing the deployer-to-provider feedback loop. Providers must actively collect and review the experiences gained from using their AI systems, utilizing various sources, including deployer feedback, incident reports, and publicly available information. The post-market monitoring system must be proportionate to the nature of the AI technology and its risks, documented in technical documentation, and integrated with the quality management system. This creates a continuous improvement obligation, where providers must utilize operational experience to enhance safety and performance.

Article 73 mandates the reporting of serious incidents and malfunctions that constitute breaches of Union law intended to protect fundamental rights. The article creates shared reporting responsibilities, stating "the provider or, where applicable, the deployer" must report incidents (Article 73(5)), acknowledging that either operator may first become aware of serious issues. Providers are obligated to report such incidents as soon as a causal connection to their system is confirmed, and in all cases, no later than 15 days from the moment they become aware of the event. This creates rapid alert mechanisms enabling authorities to respond to emerging risks and coordinate responses across member states. The reporting obligation extends to near-misses and patterns of minor incidents that collectively indicate systemic issues.

\section{The Deployer: Operational Responsibility for AI Systems}\label{deployer}

\subsection{Definition and Characteristics}\label{deployer-definition}

Article 3(4) defines the deployer as "\textit{a natural or legal person, public authority, agency or other body using an AI system under its authority except where the AI system is used in the course of a personal non-professional activity}". This definition focuses on operational use rather than development or distribution, capturing entities that make fundamental decisions about how AI systems are employed in real-world contexts where they affect individuals and society. As an operator under Article 3(8), the deployer forms the critical endpoint of the AI value chain where systems interact with affected persons, creating unique responsibilities for safeguarding fundamental rights.

The deployer definition embodies a critical regulatory insight: the risks and impacts of AI systems depend not only on their technical characteristics but fundamentally on how they are used. A facial recognition system may be technically identical whether deployed for finding missing children or conducting mass surveillance, but the deployment context transforms its risk profile and societal impact. The regulation recognizes this through Article 25(1), which can transform deployers into providers when they modify systems substantially or change their intended purpose, acknowledging that deployment decisions can fundamentally alter system characteristics.

The phrase "under its authority" establishes the key criterion distinguishing deployers from mere end-users. Authority implies decision-making power over whether, when, and how to utilize the AI system, including the ability to define usage parameters, establish operational policies, and control system application. This authority triggers the information flows mandated by Article 13, which requires providers to supply deployers with comprehensive instructions, and Article 26(5), which requires deployers to report operational issues back to providers. This captures organizations that use AI for their operations — such as banks using AI for credit decisions, hospitals utilizing diagnostic AI, and employers employing recruitment algorithms — while excluding individuals who merely interact with AI systems controlled by others. An employee required to use an AI system their employer deploys is not a deployer; the employer organization exercising authority over the system's use bears deployer obligations.

The exclusion of "personal non-professional activity" focuses regulatory attention on organizational and professional contexts where power imbalances are greatest and potential harms most significant. This exclusion covers individuals using AI for personal purposes — such as creative tools for hobbies, recommendation systems for entertainment, and personal assistants for daily tasks — while also capturing professional uses by individuals. Article 2(10) reinforces this by explicitly excluding natural persons who use AI systems for purely personal, non-professional activities from deployer obligations. At the same time, Article 49(3) requires deployers that are public authorities to register in the EU database, emphasizing institutional accountability.

\subsection{Deployer Regulations Across the Act}\label{deployer-regulations}

Article 26 outlines an extensive set of duties for deployers of high-risk AI systems, acknowledging their pivotal responsibility in ensuring the safe and compliant operation of these systems. This article contains over 20 references to deployers, making it the most deployer-centric provision in the entire regulation and establishing deployers as active participants rather than passive recipients in the operator ecosystem.

These obligations commence with the fundamental requirement in Article 26(1), which requires deployers to use high-risk AI systems in accordance with the instructions for use accompanying the system. This creates a direct link to the provider's Article 13 transparency obligations, as deployers must interpret and apply the information providers supply. However, this is not a passive obligation of blind compliance. Deployers must actively assess whether the provider's instructions are adequate for their specific use context and implement additional measures where necessary. Suppose a healthcare provider deploys a diagnostic AI system in a specialized clinical context that is not fully addressed in generic instructions. In that case, they must develop supplementary protocols to ensure safe use.

Under Article 26(2), human oversight obligations stand among the deployer's most critical duties, requiring that oversight be entrusted to natural persons with the appropriate expertise, training, and authority. This implements the human oversight measures that providers must design under Article 14(3)(b) but that deployers must operationalize, creating a shared responsibility model where providers enable and deployers execute human supervision. This creates multi-layered requirements for deployer organizations. First, they must identify individuals who are suitable for oversight roles, considering both their technical competence to understand the AI system and their domain expertise to evaluate its outputs. For neural network-based systems producing complex, non-interpretable decisions, this may require individuals with specialized training in AI explainability techniques or statistical analysis. Second, deployers must ensure that these individuals receive adequate training, not just in system operation, but also in understanding its limitations, recognizing potential failures, and knowing when to intervene. Third, and critically, oversight personnel must have genuine authority to override or halt AI system operations when necessary, requiring organizational structures that empower human judgment over automated decisions.

Article 26(3) extends certain provider-like obligations to deployers who exercise control over high-risk AI systems, particularly regarding the implementation of data governance and human oversight. This provision acknowledges that the deployer-provider boundary can blur when deployers significantly influence system operation, potentially triggering Article 25's provisions for role transformation. When deployers control the data used as inputs to AI systems, they become responsible for ensuring data quality, relevance, and representativeness for the intended purpose. This acknowledges that even well-designed AI systems can generate discriminatory or erroneous outputs if fed biased or incorrect data. A recruitment firm deploying resume-screening AI must ensure its specific applicant data doesn't introduce biases beyond those the provider addressed in the original training.

The monitoring obligations in Article 26(5) require deployers to actively observe the operation of AI systems and inform providers of any risks, serious incidents, or malfunctions that may arise. This creates the critical feedback loop that enables the provider's Article 72 post-market monitoring obligations, with deployers serving as the primary source of operational data about system performance in real-world conditions. Deployers must establish processes for detecting anomalous behaviors, unexpected outputs, or patterns suggesting degraded performance. For complex neural network systems, this may require statistical monitoring of output distributions, analysis of edge cases, or systematic testing with known inputs to ensure optimal performance. When issues are identified, deployers must follow a specific escalation chain: "immediately inform first the provider, and then the importer or distributor," creating a structured alert system through the operator hierarchy.

Article 26(6) sets forth logging requirements, obliging deployers to retain automatically generated logs from high-risk AI systems, provided those logs fall within their control and management. These logs interface with the provider's Article 19 logging design requirements and Article 73's incident reporting obligations, creating a shared evidentiary record that both operators can access for compliance and investigation purposes. The minimum retention period is six months; however, specific applications may require an extended retention period. These logs serve multiple purposes: enabling investigation of incidents, demonstrating compliance with operational requirements, and providing data for system improvement. For deployers, this requires establishing secure storage systems, implementing appropriate access controls, and developing analytical capabilities to extract insights from log data. The logs must be kept in formats enabling meaningful analysis, not merely archived as unstructured data.

Article 26(7) imposes transparency duties on deployers toward impacted individuals, obligating them to notify natural persons when using high-risk AI systems. This complements the provider's Article 50 transparency obligations but focuses on the deployment context, requiring deployers to communicate how systems are used rather than how they were designed. This notification must be clear, comprehensible, and provided before or at the time of first interaction. The requirement applies to systems listed in Annex III, with exceptions only for law enforcement uses authorized by law. Deployers must develop communication strategies tailored to their specific contexts, such as patient information for medical AI, candidate notification for recruitment systems, and citizen information for public service AI. The information provided must be meaningful, not merely a formal notification, enabling individuals to understand how AI affects decisions about them.

Article 26(8) mandates compliance with worker consultation obligations under Union and national law when AI systems affect employees. This creates a unique deployer obligation with no equivalent provider, recognizing that workplace deployment contexts raise specific concerns that require local stakeholder engagement. This acknowledges that the deployment of workplace AI raises particular concerns about surveillance, autonomy, and dignity, necessitating worker involvement in decision-making regarding its deployment. Deployers must consult with worker representatives about the introduction of AI systems, provide information about the system's purposes and impacts, and consider worker input in deployment decisions. This extends beyond mere notification to require genuine consultation that can influence deployment parameters.

Article 27 establishes one of the deployer's most significant obligations: conducting fundamental rights impact assessments for specific high-risk AI systems. Article 27(1)(d) explicitly requires deployers to consider "\textit{the information given by the provider pursuant to Article 13}", creating a direct link between provider transparency and deployer assessment obligations. This obligation applies to bodies governed by public law, private entities providing public services, and deployers of specific systems listed in Annex III, including biometric identification systems, critical infrastructure management systems, and employment-related AI systems. The assessment must identify particular risks to the rights of individuals or groups that are likely to be affected, paying specific attention to vulnerable populations that may face disproportionate impacts. Article 27(3) requires submission to market surveillance authorities, who can then coordinate with providers under Article 79 if risks are identified, creating a regulatory feedback loop through the operator chain.

Article 86 grants affected persons the right to obtain explanations directly from deployers rather than providers, recognizing that deployers control the operational context that shapes how AI decisions affect individuals. This places deployers at the interface between AI systems and fundamental rights, requiring them to translate technical system operations into meaningful explanations for affected persons.

\section{The Authorized Representative: Bridging Jurisdictional Gaps}\label{authorized-representative}

\subsection{Definition and Purpose}\label{authorized-representative-definition}

Article 3(5) defines the authorised representative as "\textit{a natural or legal person located or established in the Union who has received and accepted a written mandate from a provider of an AI system or a general-purpose AI model to, respectively, perform and carry out on its behalf the obligations and procedures established by this Regulation}". As an operator under Article 3(8), the authorised representative serves as a critical jurisdictional bridge, ensuring that non-EU providers remain accountable within the Union's regulatory framework.

The requirement for location or establishment in the Union ensures physical and legal presence, enabling effective regulatory oversight. The written mandate requirement creates formal legal relationships with clear scope and responsibilities. The representative acts as a node in the operator network, interfacing with importers who must verify their appointment (Article 23(1)(d)), distributors who may need to communicate through them (Article 24), and authorities who require a Union-based contact point (Article 22(2)).

\subsection{Authorized Representative Provisions}\label{authorized-representative-provisions}

Article 22 outlines the obligations of the authorised representative. The mandate must explicitly designate the representative and specify tasks they are authorized to perform. These obligations create parallel information channels to those of providers: representatives must maintain the same technical documentation (22(2)(a)) that importers verify (Article 23(1)(b)) and distributors check (Article 24(1)(b)). At minimum, representatives must keep technical documentation and EU declarations of conformity at authorities' disposal for ten years after market placement (22(2)(a)), provide information and documentation demonstrating conformity upon request (22(2)(b)), and cooperate with competent authorities including providing access to systems when requested (22(2)(c)).

Article 22(3) introduces a key safeguard, mandating that representatives end their mandate if they have reason to suspect that the provider is violating its obligations under the regulation. This creates a self-regulatory mechanism within the operator ecosystem, where the representative's withdrawal triggers cascade effects: importers cannot place systems on the market without valid representation (Article 23(1)(d)), effectively blocking the entire distribution chain. This creates powerful incentives for provider compliance, as losing EU representation effectively bars access to the market. The termination right also protects representatives from liability for continued non-compliance that they cannot control.

Article 54 extends the requirements for authorized representatives to providers of general-purpose AI models, placing these models on the EU market. This ensures that the operator framework encompasses both traditional AI systems and foundation models, maintaining consistent accountability structures across different AI technologies. Article 23(1)(d) requires importers to verify that non-EU providers have appointed authorized representatives before placing high-risk systems on the market. These interconnected requirements ensure comprehensive coverage of non-EU actors in the AI value chain.

\section{The Importer: First Line of Defense for Market Entry}\label{importer}

\subsection{Definition and Market Role}\label{importer-definition}

Article 3(6) defines the importer as "\textit{a natural or legal person located or established in the Union that places on the market an AI system that bears the name or trademark of a natural or legal person established in a third country}". As an operator positioned at the crucial junction between external providers and the Union market, importers serve as gatekeepers who must verify compliance before systems enter the regulatory perimeter.

The reference to systems bearing third-country names or trademarks clarifies that the decisive factor is the non-EU origin of the system's provider, not necessarily where the system was manufactured or developed. This definition ties with Article 25(1), which can transform importers into providers if they register their name in the system or make substantial modifications, emphasizing the fluidity of operator roles based on actual conduct rather than formal designations.

\subsection{Importer Obligations Throughout the Regulation}\label{importer-obligations}

Article 23 establishes comprehensive importer obligations that go far beyond simple customs clearance. These obligations create a verification cascade: importers must confirm provider compliance (23(1)(a)), which providers must demonstrate through documentation (Article 11), which deployers will later rely upon (Article 26(9)). Before placing high-risk AI systems on the market, importers must ensure the appropriate conformity assessment has been carried out (23(1)(a)), technical documentation has been drawn up (23(1)(b)), the system bears required CE marking and is accompanied by EU declaration of conformity and instructions for use (23(1)(c)), and the provider has appointed an authorized representative (23(1)(d)).

Article 23(2) requires importers who have reason to consider an AI system non-compliant to refrain from placing it on the market until it is brought into conformity with the applicable requirements. Where systems present risks, importers must "immediately inform the provider of the system and the market surveillance authorities", creating dual reporting channels that ensure both technical remediation (through the provider) and regulatory awareness (through authorities). This makes active verification obligations rather than passive acceptance of provider declarations.

Article 23(3) mandates that importers ensure storage and transport conditions do not jeopardize compliance. At the same time, Article 23(4) requires them to indicate their name, registered trade name or trademark, and contact address on the system or its documentation. This identification requirement ensures traceability through the operator chain, enabling deployers to identify responsible parties and authorities to map distribution networks. Article 23(5) requires importers to retain copies of EU declarations of conformity for ten years and ensure that technical documentation remains accessible to the relevant authorities.

Article 23(6) extends provider-like obligations to importers for compliance verification, requiring them to perform testing when they have reason to believe systems are non-compliant. This creates a secondary compliance check that supplements provider obligations, with Article 79(2) requiring importers as "relevant operators" to cooperate with authorities during risk investigations. Article 23(7) mandates cooperation with competent authorities, while Article 23(8) requires immediate corrective action for non-compliant systems they have placed on the market.

\section{The Distributor: Maintaining Compliance Through Distribution}\label{distributor}

\subsection{Definition and Supply Chain Position}\label{distributor-definition}

Article 3(7) defines the distributor as "a natural or legal person in the supply chain, other than the provider or the importer, that makes an AI system available on the Union market". As operators positioned throughout the distribution network, distributors maintain the integrity of AI systems as they move from initial market placement to final deployment, serving as compliance checkpoints throughout the supply chain.

The phrase "other than the provider or the importer" clarifies that these roles are mutually exclusive---an entity cannot simultaneously be a distributor and a provider or importer for the same system. However, Article 25(1) can transform distributors into providers if they modify systems or change their intended purpose, demonstrating the dynamic nature of operator categorization based on actual activities rather than initial roles.

\subsection{Distributor Regulations}\label{distributor-regulations}

Article 24 establishes distributor obligations calibrated to their more limited role, while recognizing their importance in maintaining compliance throughout the distribution chain. These obligations create a verification relay: distributors check what importers have verified (Article 23) and what providers have produced (Articles 11, 13), ensuring multiple compliance checkpoints before systems reach deployers. Before making high-risk AI systems available, distributors must verify that the system bears CE marking (24(1)(a)), is accompanied by required documentation and instructions (24(1)(b)), that provider and importer have complied with their obligations regarding identification and documentation (24(1)(c)).

Article 24(2) requires distributors who consider systems non-compliant to refrain from making them available until they are brought into conformity, paralleling the obligations of importers. When distributors identify non-compliance, they must "inform the provider or the importer," creating upstream feedback that can trigger the provider's Article 20 corrective action obligations. Article 24(3) mandates ensuring storage and transport conditions maintain compliance, while Article 24(4) requires cooperation with authorities and upstream actors to ensure conformity.

Article 24(5) requires distributors to provide authorities with information and documentation demonstrating conformity, while Article 24(6) mandates immediate corrective action for non-compliant systems they have made available. These obligations ensure distributors actively participate in the Article 79 risk management procedures and Article 82 responses to compliant but risky systems, functioning as full operators rather than passive intermediaries.

\section{The Operator Concept: Unifying Framework for AI Ecosystem Governance}\label{operator-concept}

\subsection{Definition and Critical Assessment}\label{operator-definition-critical}

Article 3(8) describes an "operator" as encompassing a provider, product manufacturer, deployer, authorized representative, importer, or distributor, serving as a unifying term for all key participants in the AI ecosystem. However, this collective definition introduces significant interpretative challenges and potential confusion. Rather than clarifying the regulatory framework, the operator concept risks obscuring the distinct roles and responsibilities of each actor in the AI value chain.

From a critical perspective, the introduction of this umbrella term is problematic for several reasons. First, it creates semantic ambiguity: when the regulation refers to "operators," it becomes unclear whether specific obligations apply equally to all categories or require differentiated implementation based on their actual role. This ambiguity could lead to legal uncertainty and inconsistent application across Member States. Second, the operator concept may inadvertently dilute accountability by allowing entities to invoke the general "operator" status rather than acknowledging their specific role-based obligations. Third, it complicates compliance efforts, as organizations must parse whether each "operator" reference applies to their particular situation or represents a general principle.

The operator definition is a legislative convenience rather than a substantive regulatory innovation. While it enables more concise drafting in specific articles, this efficiency comes at the cost of precision. The regulation could have achieved the same objectives by explicitly listing applicable roles in each provision, which would enhance clarity even if it increased textual length. The fact that many articles still specify particular operator types (e.g., provider, deployer) suggests that the unified concept fails to eliminate the need for role-specific provisions.

\subsection{Problematic Implementation Throughout the Regulation}\label{operator-implementation-problems}

The inclusion of "product manufacturer" addresses scenarios where AI systems are safety components of products covered by Union harmonization legislation. Article 25(3) automatically designates product manufacturers as providers when they integrate high-risk AI systems into their products, ensuring no regulatory gaps exist between AI-specific and product safety regulations. Article 2(1)(b) expands the regulation's coverage to encompass providers that market or deploy AI systems serving as safety components of products regulated under the legislation listed in Annex I. However, this inclusion within the operator framework further muddies the waters, as product manufacturers have fundamentally different business models and regulatory contexts than pure AI providers or deployers.

The implementation of the operator concept throughout the regulation reveals its limitations. Article 79 demonstrates both the utility and confusion of the operator approach: while paragraph 2 requires all "relevant operators shall cooperate" with authorities, the use of "relevant" immediately raises questions about which operators are relevant in which circumstances. The regulation provides no clear criteria for determining relevance, leaving this critical determination to case-by-case interpretation. Article 83 stipulates that operators must provide authorities with all necessary documentation and information to demonstrate conformity with the relevant regulations. Article 84 grants authorities the power to require information from any operator. These provisions, while establishing baseline obligations, fail to account for the vastly different capacities and access to information that different operator types possess. A distributor, for instance, may have limited ability to provide technical documentation that remains primarily under the control of the provider.

Article 99 on penalties nominally structures sanctions around the operator framework, with paragraph 1 requiring member states to establish "penalties...applicable to infringements of this Regulation by operators". However, paragraph 4 immediately abandons the unified approach by differentiating penalty levels based on specific operator types, implicitly acknowledging that the operator concept lacks sufficient granularity for liability allocation. The maximum fines vary significantly: violations of prohibited practices can reach €35 million or 7\% of worldwide annual turnover, while violations of other operator obligations face lower ceilings. This differentiation undermines the coherence of the operator concept---if all operators were truly equivalent for regulatory purposes, why would penalties require such careful differentiation?

The operator concept also complicates the dynamic role transformations under Article 25. When a deployer substantially modifies a high-risk AI system (Article 25(1)(b)) or changes its intended purpose (Article 25(1)(c)), they become a provider with full provider obligations. However, referring to both the original and transformed entity as "operators" obscures the fundamental shift in responsibilities and capabilities required. A deployer-turned-provider must suddenly develop technical documentation, quality management systems, and conformity assessment procedures---capacities that the operator framework fails to distinguish or address.

Article 94 extends procedural rights to all operators in proceedings concerning general-purpose AI models; however, this blanket approach overlooks the vastly different stakes and capabilities that other operators have in such proceedings. A small distributor has fundamentally different interests and expertise from a major provider, yet the operator framework treats them as equivalent parties. Article 95 enables operators to develop codes of conduct individually or collectively. Still, the practical challenges of creating codes that meaningfully address the disparate concerns of providers, deployers, importers, and distributors suggest that the operator concept creates false equivalencies.

\subsection{Consequences and Recommendations}\label{operator-consequences}

The operator framework, while perhaps intended to create a sophisticated governance ecosystem, instead risks creating confusion and compliance challenges:

\begin{itemize}
\item Information flows become ambiguous when "operator" obligations don't specify which role should initiate or receive information
\item Verification responsibilities blur when multiple "operators" have overlapping but undefined duties
\item Role transformations under Article 25 become more complex when entities must navigate both their specific role and general operator obligations
\item Cooperation mandates lack specificity when Articles 20, 79, and 82 refer to "operators" without clarifying coordination mechanisms
\item Penalties lose proportionality when the operator framework obscures actual control and capability differences
\end{itemize}

From a critical standpoint, the operator concept represents a missed opportunity for regulatory clarity. The AI Act would have been better served by:

\begin{enumerate}
\item Eliminating the operator definition and using specific role references throughout
\item Creating clear hierarchies of responsibility based on actual control rather than formal categories
\item Developing role-specific compliance frameworks that acknowledge different capabilities and market positions
\item Establishing explicit coordination mechanisms between specific roles rather than vague operator cooperation duties
\end{enumerate}

The operator concept ultimately demonstrates a tension between legislative efficiency and regulatory precision. While it may simplify specific drafting challenges, it complicates the interpretation, implementation, and enforcement of these regulations. For a regulation governing complex and rapidly evolving technology, such ambiguity represents a significant weakness that could undermine the effectiveness and legal certainty of the AI Act. Future amendments should consider dismantling the operator framework in favor of explicit, role-specific provisions that better reflect the realities of the AI value chain.

\begin{figure}[H]
\centering
\includegraphics[width=0.9\textwidth]{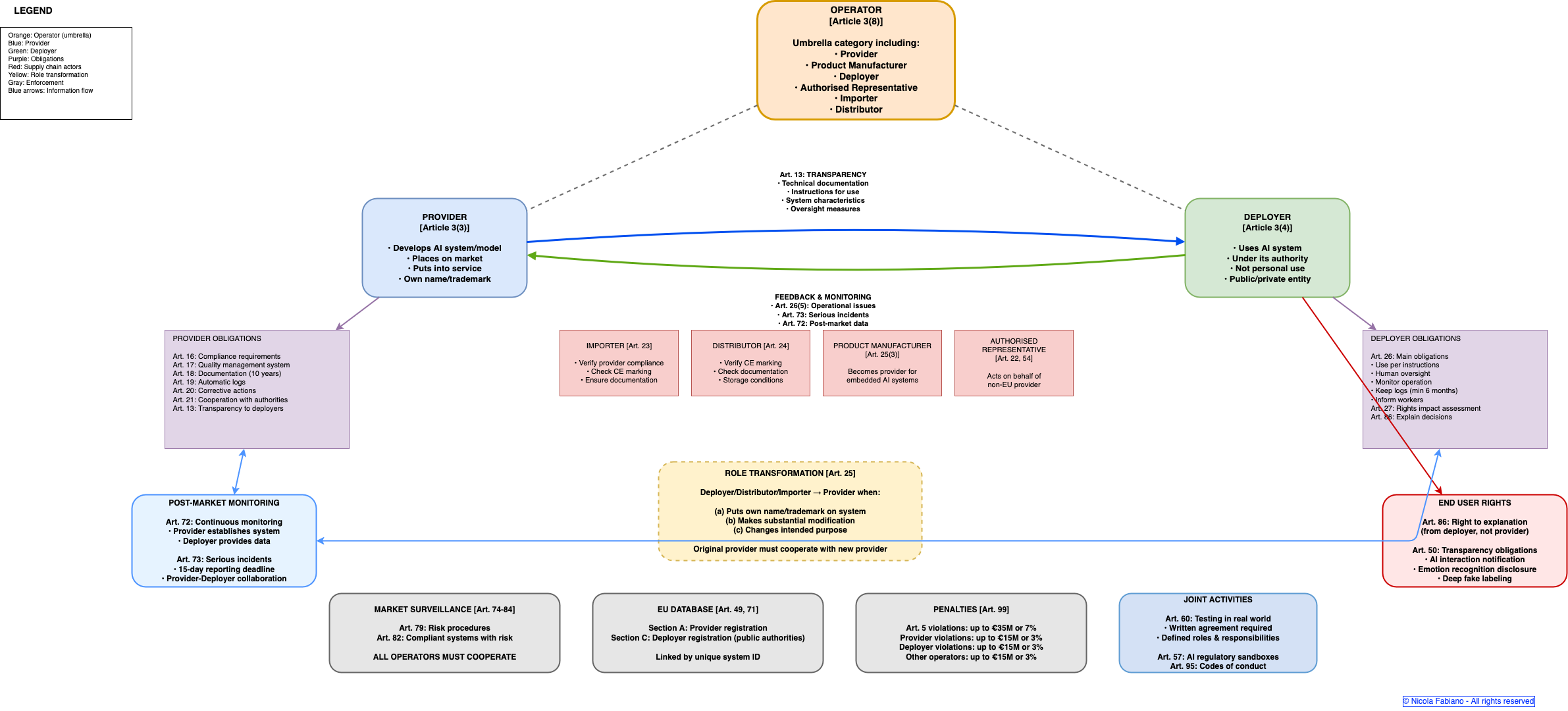}
\caption{Subjects Interactions}
\label{fig:subjects-interactions}
\end{figure}
\section{Role Transformations and Dynamic Responsibilities}\label{role-transformations}

Article 25 establishes critical mechanisms whereby operators assume different roles based on their actions, ensuring accountability follows control regardless of formal designations. That represents one of the regulation's most sophisticated elements, preventing circumvention through creative structuring.

Article 25(1) specifies three scenarios where distributors, importers, deployers, or other third parties assume provider obligations: (a) placing an AI system on the market or putting it into service under their name or trademark; (b) making substantial modifications to high-risk systems already on the market; (c) modifying the intended purpose of non-high-risk systems making them high-risk.

These transformations have profound implications. When a distributor rebrands an AI system, they assume full provider obligations, including conformity assessment, technical documentation, and post-market monitoring. When a deployer substantially modifies a high-risk system, they become responsible for ensuring continued compliance with relevant regulations. These mechanisms ensure that entities exercising provider-like control bear commensurate responsibilities.

Article 28 establishes that product manufacturers placing products on the market or putting them into service with embedded high-risk AI systems become providers of those AI systems. They must ensure the AI system complies with this regulation and integrate conformity assessment into existing procedures under relevant sectoral legislation.

\section{Information Flows and Cooperation Requirements}\label{information-flows}

The regulation establishes comprehensive information flows between operators, ensuring effective governance across distributed value chains. These flows operate in multiple directions, creating networks of accountability and support.

Downstream information flows run from providers through the supply chain to end-users, also known as deployers. Article 13 requires providers to ensure that high-risk systems are accompanied by instructions for use that contain the information specified in Annex IV. Article 53(1)(b) requires general-purpose AI model providers to provide downstream providers with technical documentation enabling integration into AI systems. Importers and distributors must verify that this information is present and pass it along the chain.

Upstream feedback flows enable continuous improvement. Article 26(5) requires deployers to monitor operations and inform providers and distributors about risks and incidents. Article 72 requires providers to establish post-market monitoring systems that collect feedback from deployers. Article 73 mandates the reporting of serious incidents, creating rapid alert mechanisms for safety issues.

Lateral information sharing between authorities ensures coordinated oversight. Article 74 requires market surveillance authorities to share information about non-compliant systems. Article 88 establishes the EU database for high-risk AI systems, providing authorities with access to registration information. These mechanisms prevent regulatory arbitrage and ensure consistent enforcement of regulations.

\section{Enforcement and Compliance Mechanisms}\label{enforcement}

The regulation establishes differentiated enforcement mechanisms recognizing operators' varying roles and capabilities. Article 99 establishes administrative fines up to €35 million or 7\% of worldwide annual turnover for violations of prohibited practices, up to €15 million or 3\% for violations of operator obligations, and up to €7.5 million or 1.5\% for supplying incorrect information.

Market surveillance authorities designated under Article 74 have the power to require information from any operator (Article 84), access data and documentation (Article 85), and conduct both physical and remote inspections (Article 86). They can require operators to take corrective action, withdraw or recall systems, and prohibit or restrict market availability (Article 87).

For high-risk AI systems, conformity assessment procedures vary depending on the role of the operator. Providers must undergo an assessment before being placed on the market, while importers must verify that the evaluation has been completed. Distributors must confirm the presence of CE marking and accompanying documentation. Deployers are not subject to conformity assessment but must conduct impact assessments for specific applications.

\section{Temporal Implementation and Transitional Provisions}\label{temporal-implementation}

The provisions of the AI Act analyzed in this document come into force on different dates, creating a phased implementation approach that organizations must carefully navigate.

The regulation's phased implementation recognizes different preparation requirements for various operators. Article 113 establishes staggered application dates: prohibited practices take effect on February 2, 2025; general-purpose AI model rules take effect on August 2, 2025; most provisions, including operator obligations, take effect on August 2, 2026; and requirements for specific pre-existing systems take effect on August 2, 2027.

These timelines establish distinct compliance horizons for various operators.

As of August 2, 2025, approximately 28.3\% of the AI Act's provisions have become applicable, including crucial governance structures and obligations for general-purpose AI models.

The provider obligations under Articles 16-21 and deployer obligations under Articles 26-27, which form the core of the operational framework discussed above, will become fully enforceable on August 2, 2026, when the majority of the Act's provisions take effect. Notably, the transformation mechanisms under Article 25, which can convert deployers or distributors into providers, also activate on this date. The authorization and market surveillance framework, including the obligations of importers (Article 23) and distributors (Article 24), similarly becomes operational on August 2, 2026.

Organizations should note that Article 6(1) regarding classification rules for high-risk AI systems has a deferred implementation until August 2, 2027, providing additional time for compliance with classification requirements.

The sanctions regime under Article 99 becomes enforceable on August 2, 2026, except for Article 101, which concerns financial penalties for general-purpose AI model providers and has a separate timeline.

Article 111 provides transitional provisions for AI systems already on the market, provided they undergo no substantial modifications.

This staggered approach enables organizations to progressively adapt their compliance frameworks, ensuring that the most critical safety and governance provisions receive immediate attention.

\section{International Dimensions and Extraterritorial Application}\label{international-dimensions}

The operator definitions interact with the regulation's extraterritorial reach, as outlined in Article 2. The regulation applies to providers and deployers established outside the Union, provided that the output produced is used within the Union. That necessitates a mechanism for authorized representatives to ensure EU-based accountability for non-EU providers.

The importer definition establishes the first EU-established operator in many supply chains, assuming special responsibility for ensuring that third-country systems meet EU standards. Importers must navigate the differences between EU requirements and those of their country of origin, which may require system modifications or additional documentation.

Mutual recognition agreements may eventually allow third-country conformity assessments to meet EU requirements, but until then, importers bear full responsibility for verification. The regulation's influence on global AI governance partly operates through these importer obligations, as third-country providers must meet EU standards to access the market through any importer.

\section{Conclusions}\label{conclusions}

This comprehensive analysis of the operator categories defined in Article 3 of the AI Act reveals a sophisticated regulatory architecture that distributes responsibilities across the AI value chain while maintaining clear accountability. The six operator types — providers, deployers, authorized representatives, importers, distributors, and product manufacturers — each bear obligations tailored to their role and level of control over AI systems.

The provider definition establishes primary responsibility with those who develop and place AI systems on the market, supported by extensive obligations throughout the regulation, from quality management to post-market monitoring. The deployer definition captures operational responsibility, recognizing that deployment context significantly affects risk. Authorized representatives ensure accountability for non-EU providers, while importers and distributors serve as gatekeepers, maintaining compliance through distribution chains.

The operator concept unifies these actors under common provisions while preserving role-specific obligations. Dynamic transformation mechanisms ensure accountability follows control, preventing circumvention through corporate structuring. Comprehensive information flows and cooperation requirements create networks enabling effective governance across complex value chains.

The regulation's treatment of these defined operators demonstrates a sophisticated understanding of modern AI ecosystems. Rather than imposing uniform obligations, it creates differentiated requirements reflecting each operator's position and capabilities. This approach strikes a balance between comprehensive coverage and practical feasibility, ensuring effective governance without imposing disproportionate burdens.

Success in implementing this framework requires operators to understand not just their primary role but also how their actions might trigger additional obligations. Organizations must establish systems for managing their operators' obligations while coordinating with other operators in their value chains. Authorities must develop the capabilities to oversee these complex, multi-actor systems while ensuring consistent enforcement of regulations.

As AI technology continues evolving, the operator framework provides a stable foundation for governance. The definitions are sufficiently broad to encompass new business models and technical architectures while specific enough to provide legal certainty. The regulation's built-in adaptation mechanisms allow refinement of operator obligations without requiring fundamental restructuring.

The AI Act's operator framework represents a significant evolution in technology governance, moving beyond simple manufacturer liability to encompass all actors influencing AI system development and deployment. This comprehensive approach, grounded in the precise definitions outlined in Article 3 and elaborated throughout the regulation's 113 articles, establishes Europe as a leader in AI governance while providing a model that is likely to influence global regulatory development.

\section{References}\label{references}

Adams Jon (2024), AI Foundations of Neural Networks: Easy To Read Guide Introducing the Foundations Of Neural Networks and AI, Green Mountain Computing, 1st edition (February 4, 2024)

Bridgelall R., Unraveling the mysteries of AI chatbots, Artif Intell Rev 57, 89 (2024) - \url{https://doi.org/10.1007/s10462-024-10720-7}.

European Commission, Digital Single Market, \url{https://ec.europa.eu/digital-single-market/en/news/communication-artificial-intelligence-europe}

European Commission - Digital Strategy - \url{https://digital-strategy.ec.europa.eu/en/library/ethics-guidelines-trustworthy-aiStuart}

European Union Eur-Lex (2024), Regulation (EU) 2024/1689 of the European Parliament and of the Council of 13 June 2024 laying down harmonised rules on artificial intelligence and amending Regulations (EC) No 300/2008, (EU) No 167/2013, (EU) No 168/2013, (EU) 2018/858, (EU) 2018/1139 and (EU) 2019/2144 and Directives 2014/90/EU, (EU) 2016/797 and (EU) 2020/1828 (Artificial Intelligence Act) - \url{https://eur-lex.europa.eu/legal-content/EN/TXT/PDF/?uri=OJ:L_202401689}

Fabiano Nicola (2025), Artificial Intelligence, Neural Networks and Privacy: Striking a Balance between Innovation, Knowledge, and Ethics in the Digital Age, Fireze, goWare, ISBN: 9788833636849 - \url{https://www.goware-apps.com/libri/artificial-intelligence-neural-networks-and-privacy/}

Fabiano Nicola (2020), GDPR \& Privacy: awareness and opportunities. The approach with the Data Protection and Privacy Relationships Model (DAPPREMO), Firenze, goWare, ISBN: 9788833634043 - \url{https://www.goware-apps.com/libri/gdpr-privacy-awareness-and-opportunities-the-approach-with-the-data-protection-and-privacy-relationships-model-dappremo-nicola-fabiano/}

Farisco Michele (2024), Evers Kathinka, Changeux Jean-Pierre, Is artificial consciousness achievable? Lessons from the human brain, Neural Networks, Volume 180, 2024, 106714, ISSN 0893-6080, \url{https://doi.org/10.1016/j.neunet.2024.106714} .

OECD (2024), Recommendation of the Council on Artificial Intelligence - \url{https://legalinstruments.oecd.org/en/instruments/oecd-legal-0449}

Russell Stuart - Peter Norvig (2020), "Artificial Intelligence: A Modern Approach, 4th edition", Published by Pearson (April 28, 2020) © 2021 - \url{https://www.pearson.com/en-us/subject-catalog/p/artificial-intelligence-a-modern-approach/P200000003500}.

Stanford University, Artificial Intelligence - \url{https://setr.stanford.edu/technology/artificial-intelligence/2025}

Stanislav Ivanov, Craig Webster (2024), Automated decision-making: Hoteliers' perceptions, in Technology in Society, Volume 76, 2024, 102430, ISSN 0160-791X - \url{https://doi.org/10.1016/j.techsoc.2023.102430}.

\end{document}